# Software Defined Demodulation of Multiple Frequency Shift Keying with Dense Neural Network for Weak Signal Communications

Mykola Kozlenko
*Department of Information Technology*
*Vasyl Stefanyk Precarpathian National University*
Ivano-Frankivsk, Ukraine
https://orcid.org/0000-0002-2502-2447

Vira Vialkova
*Department of Cyber Security and Information Protection*
*Taras Shevchenko National University of Kyiv*
Kyiv, Ukraine
https://orcid.org/0000-0001-9109-0280

*Abstract*—In this paper we present the symbol and bit error rate performance of the weak signal digital communications system. We investigate orthogonal multiple frequency shift keying modulation scheme with supervised machine learning demodulation approach using simple dense end-to-end artificial neural network. We focus on the interference immunity over an additive white Gaussian noise with average signal-to-noise ratios from -20 dB to 0 dB.

## I. Introduction

Weak signals are very common in various long-distance digital radio communications, such as Earth-Moon-Earth (EME), meteor scatter, VHF-DX, and other. EME communications, also known as the Moon Bounce, is a communication technology based on the reflection of electromagnetic waves from the surface of the Moon. Nowadays it is used by community of amateur radio operators [1], [2]. 2 meter, 70-centimeter, and 23-centimeter bands are commonly used for EME communications. Widely used modulation modes are continuous wave keying (CW) with Morse code, digital protocol JT65, and voice with Single Side Band (SSB) [3], [4]. The main challenges are the following: very low signal-to-noise ratio (SNR), echo delay, time spread, intersymbol interference, Doppler Effect, libration fading, and polarization effects [4]. It is very difficult to detect and extract weak signals. One of the traditional approaches for obtaining stable communication link in such conditions is based on the use of complex signals, in particular, spread spectrum signals and methods [5], [6]. Frequency hopping spread spectrum (FHSS), direct-sequence spread spectrum (DSSS), time-hopping spread spectrum (THSS), chirp spread spectrum (CSS), and combinations of these techniques are commonly used as basic methods for signal generating and processing [5]. Also, stochastic communication systems based on various techniques of noise shift keying have good performance [7]. Particularly, in [8] the system is presented where the standard deviation of the band-limited Gaussian noise is used for representation of binary symbols and the entropy estimation of the received signal is used for demodulation and detection. In this paper we consider JT65A digital communication protocol. It has very good interference immunity and possibility to work at extremely low signal-to-noise ratio. We investigate the possibility of successful using of supervised machine learning approach based on artificial dense neural network for signal demodulation.

## II. Previous Works

In the paper [9] the exceptional performance of learning-based gain is presented. It is based on the use of deep convolutional neural network for demodulation of a Rayleigh-faded wireless data signal. In the work [10] it is shown the use of existing deep learning (DL) methods DBN and SAE for signal demodulation in short range multi-path channel. Reference [11] presents an investigation of the design and implementation of machine learning (ML) based demodulation methods in the physical layer of visible light communications (VLC) systems. In the paper [12] a neural network-based method to demodulate digital signals is presented. After training with different modulation schemes, the learning-based receiver can perform demodulation without changing receiver hardware by loading certain parameters based on the modulation scheme. In the work [13] next-generation wireless networks are presented. These networks are expected to support extremely high data rates and radically new applications using machine learning as one of the most promising artificial intelligence tools, conceived to support smart radio terminals. In the work [14] several novel applications of DL for the physical layer are presented and discussed. By interpreting a communications system as an autoencoder, the authors developed a fundamental new way to think about communications system design as an end-to-end reconstruction task. In the work [15] the use of such approaches is declared for creation of pseudo satellite radio navigation system. The paper [16] is about the use of spread spectrum signals and statistical entropy demodulation in telecommunication system of household power supply meters. In the letter [17] demodulation method based on an artificial neural network is proposed. It is used for demodulation of optical eigenvalue modulated signals using on-off encoding.

## III. Data Synthesis

Supervised machine learning approach requires a large amount of training data. Thus, in this research we used artificially synthesized data. The use of artificial data has certain limitations and additional research is required to prove the feasibility of this model for real-channel wideband, narrowband, and pulse interferences with more dynamic behaviour. The training set contains 100000 m-ary frequency shift keying (MFSK) signal fragments according to JT65A communication protocol. JT65A is a radio communication protocol developed by Joe Taylor [18]. Information symbols are represented by one of 64 predefined orthogonal frequencies and the synchronizing tone at 1270.5 Hz. The exact frequency value can be calculated by its ordinal number *m* as follows:

$$f = 1270.5 + 2.6817 \cdot (m+2), \quad (1)$$

$$m = 0,1,2,\ldots,63$$

The duration of each symbol interval is 0.3715 second. It consists of 4096 samples. Sample rate is 11025 samples per second. The bandwidth is limited to 2500 Hz. Each symbol carries exactly 6 information bits. The entire transmission contains 126 contiguous time intervals and lasts for 46.8 seconds. While synthesizing the dataset, the sinusoidal wave with random phase was mixed with additive white Gaussian noise (AWGN). SNR was chosen randomly in the range from -20 dB to 0 dB. The example of artificially synthesized signal waveform in time domain at SNR = -10 dB is shown in Fig. 1.

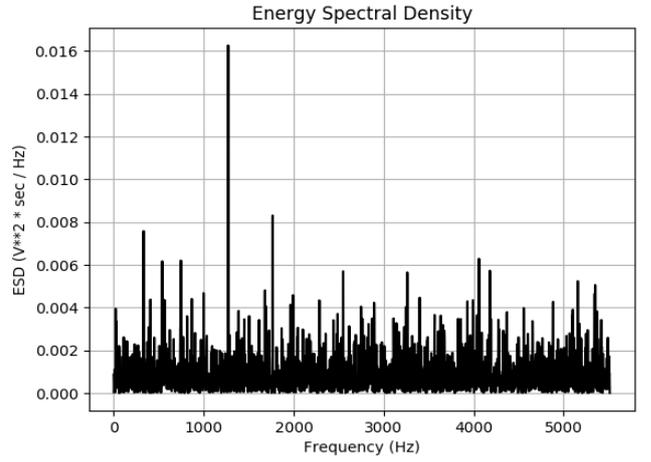

Fig. 2. ESD of synthesized JT65A symbol signal at SNR = -20 dB.

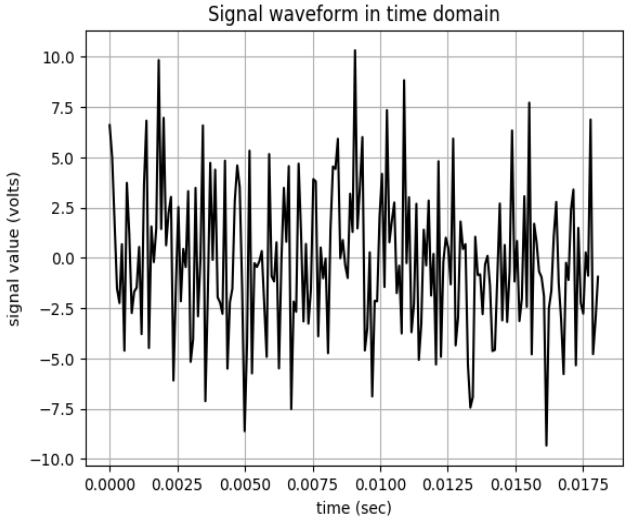

Fig. 1. Signal waveform in time domain (first 200 samples) at SNR = -10 dB.

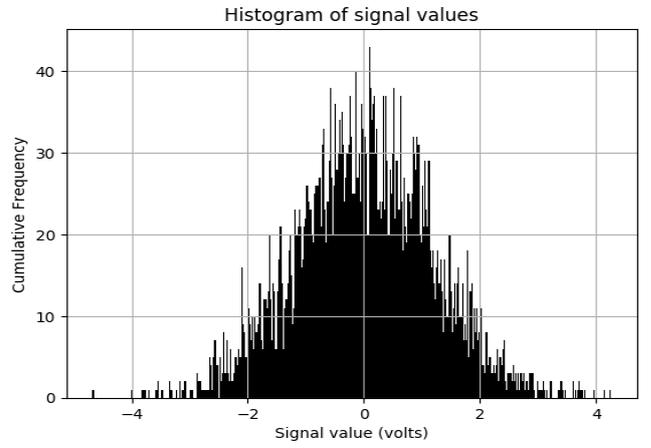

Fig. 3. Histogram of one symbol signal at SNR = -10 dB.

Energy Spectral Density (ESD) of synthesized signal was obtained using the Fast Fourier Transform (FFT) [19]. The FFT (2) was computed using numpy.fft Python library module.

$$S_x(k \cdot F) = T_s \cdot \sum_{n=0}^{N-1} x(n \cdot T_s) \exp(-j2\pi nk/N), \quad (2)$$

where $S_x(kF)$ - discrete amplitude density, V/Hz

$x(nT_s)$ - discrete signal value, V

$N$ - number of signal samples

$T_s$ - sampling time interval, $T_s = 1/f_s$, sec

$f_s$ - sampling frequency,

$k$ - frequency index, $k = 0,1,\ldots,N-1$

$n$ - time index, $n = 0,1,\ldots,N-1$.

Energy spectral density was computed at frequencies $kF$ as $|S_x(kF)|^2$, with interval $F = 1/NT_s$ in frequency domain (Fig. 2). Signal distribution is shown in the Fig. 3.

## IV. NEURAL NETWORK MODEL DESIGN

It is a good practice to start with a very simple model at first. Our baseline model was a Softmax classifier based on "brute-force" end-to-end dense deep neural network [20] (Fig. 4, 5, 6). The used tools were: TensorFlow 1.15.0 [21], Keras 2.2.4-tf [22], numpy 1.17.4 [23], [24], pandas 0.25.3 [25], Python 3.6.8 [26], [27]. The Tensorboard web application was used for visualization of training scalars and neural network structures.

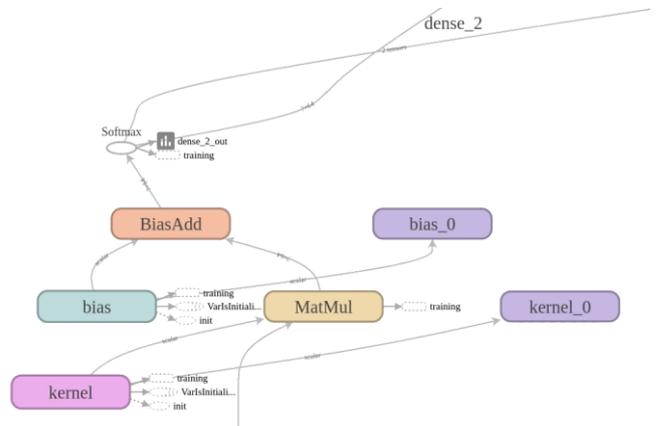

Fig. 4. Structure of one dense layer with the softmax activation function.

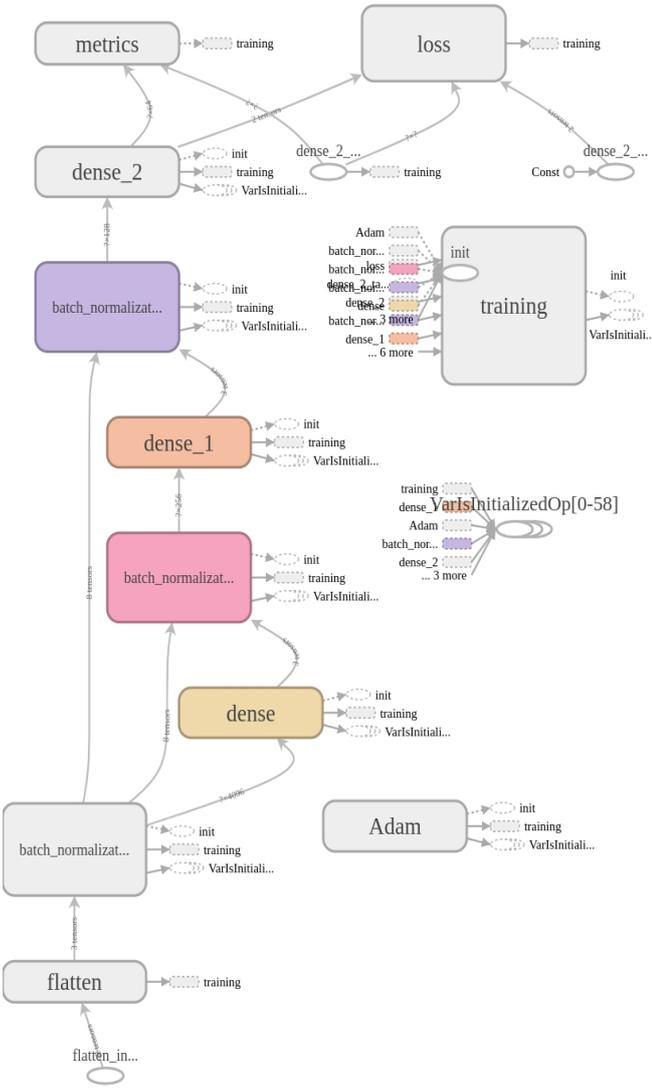

Fig. 5. Structure of the dense artificial neural network model.

The model has totally 1107904 parameters, among them 1098944 trainable parameters, and 8960 non-trainable.

Architecture details and hyper-parameters selection:

*A. Number of Layers*
- There are one input, one output, and two hidden layers in the design.

*B. Number of neuron units per layer*
- Input layer: 4096 (number of samples per signal symbol interval), Output layer: 64 (number of possible information symbol values), 1-st hidden fully connected layer: 256, 2-nd hidden fully connected layer: 128.

*C. Activation function*
- ReLU [28], [29] at the hidden layers, Softmax at the Output layer.

*D. Optimizer*
- Adam [30], [31]. (Keras built-in).

*E. Number of epochs*
- 5.

*F. Loss Function*
- Categorical cross-entropy (3).

$$H(p,q) = -\sum_{x \in X} p(x) \cdot \log(q(x)) \quad (3)$$

*G. Metrics*
- Error rate, accuracy, precision, recall.

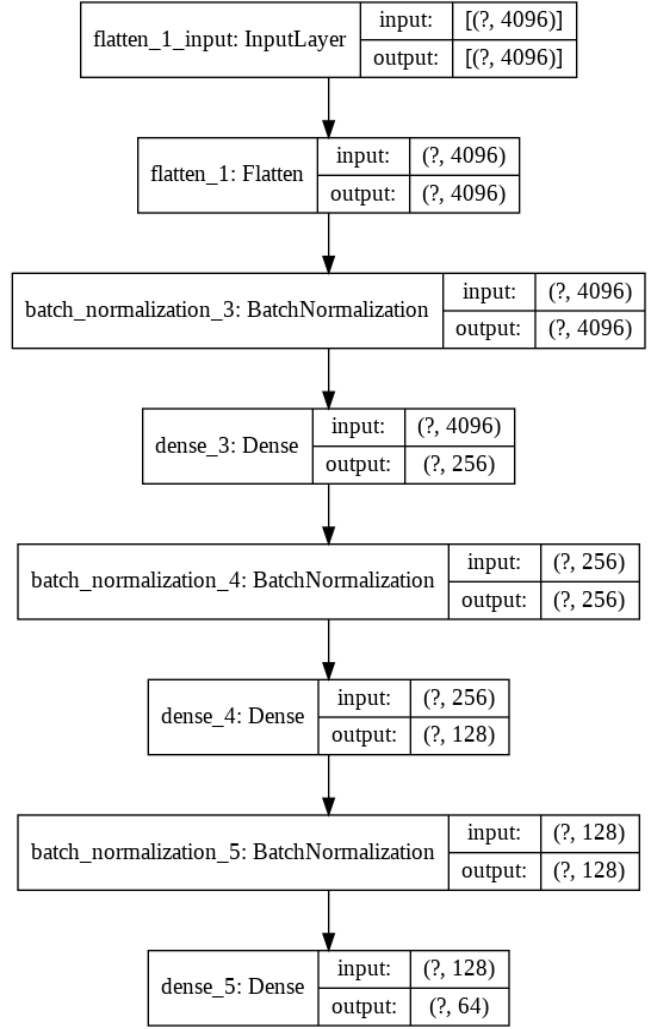

Fig. 6. Dimentionality of data passing through the model.

## V. TRAINING AND EVALUATING

Training of the model was performed with the training dataset using Google Colaboratory. Colaboratory is a free Jupyter notebook environment that does not require any setup and runs entirely in the cloud. With Colaboratory one can write and execute Python code, save and share analyses, and access graphics processing unit (GPU) as well as tensor processing unit (TPU) from browser [32]. Training procedure takes approximately 260 microseconds per one step (26 seconds per epoch) with GPU hardware accelerator.

The batch size (the number of samples per gradient update) is 32. Training loss and accuracy against epoch number and step number are shown in Fig. 7, 8 correspondently. Values are taken at the end of an epoch or step.

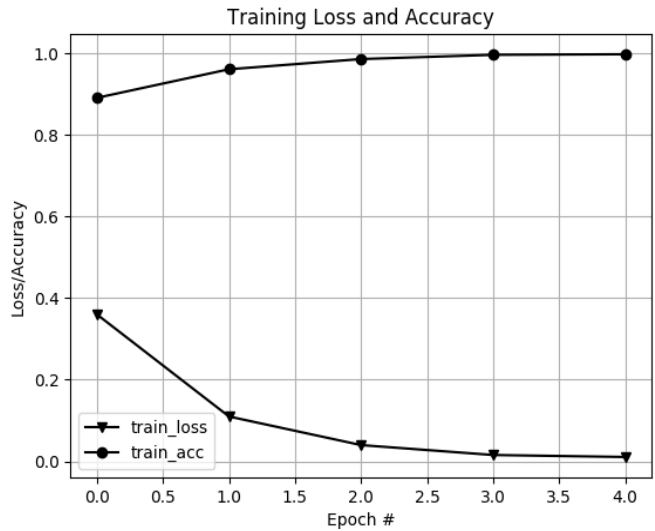

Fig. 7. Training loss and accuracy against epoch number.

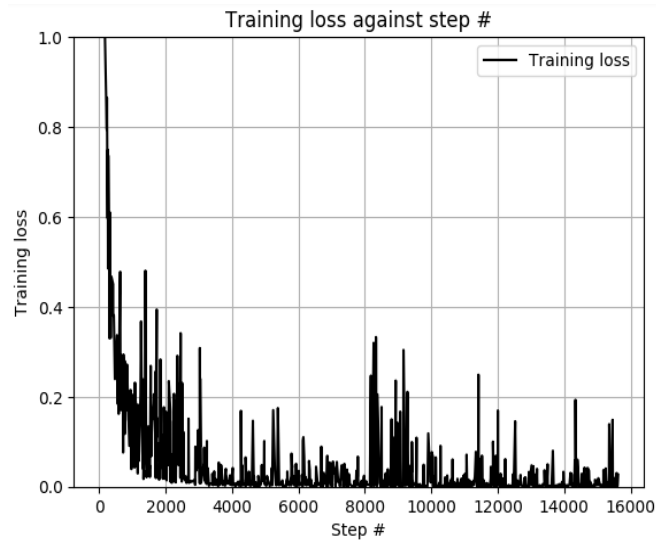

Fig. 8. Training loss against step number.

We used post-predict evaluation in order to obtain the quality metrics. Test set went through the prediction method. Test set contained 10000 symbol signals for each needed signal-to-noise ratio. After that, predictions were compared to the ground truth and the confusion matrix was derived. The following class-wise and macro/micro-averaged metrics were received from the confusion matrix: error rate, accuracy, true positive rate (recall), positive predictive value (precision).

## VI. COMPUTATIONAL COMPLEXITY OF INFERENCE PHASE

Complexity is the first priority issue for real time signal processing. Finding the run-time complexity of the forward propagation procedure can be done by measurement time of one sample processing on various hardware platforms.

TABLE I. RUN-TIME COMPLEXITY

| Hardware platform | Prediction Time (us/sample) |
|---|---|
| NVIDIA TESLA K80 | 34 |
| Intel i5-3470 | 70 |
| Raspberry Pi 4 | 85 |
| Samsung Galaxy J2 2018 | 80 |

This can clarify the feasibility of the approach for various systems. Averaged results are presented in the Table I. The processing time of one symbol does not exceed the duration of the symbol interval. As it has been found out, it is possible to perform real-time JT65A protocol demodulation with any of mentioned modern hardware platforms.

## VII. RESULTS

Common used metric for the quality of any digital communications system is a dependence of bit error rate against normalized signal-to-noise ratio ($E_b/N_0$).

Plot of symbol error rate against SNR is shown in the Fig. 9.

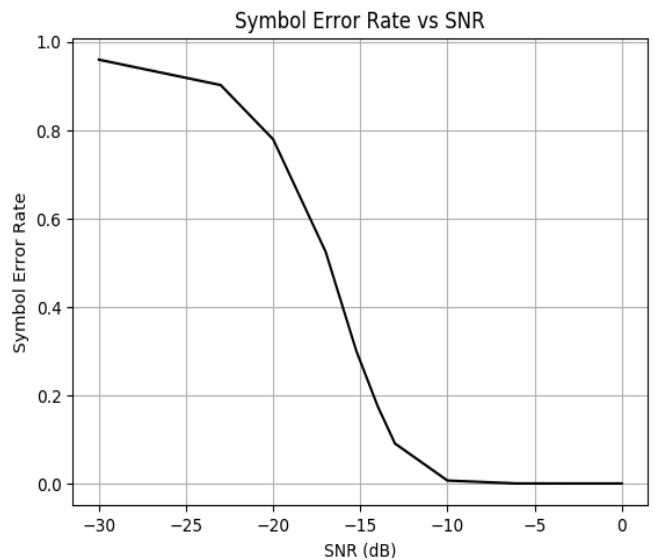

Fig. 9. Demodulation Symbol Error Rate against SNR.

One of the main equations [5] of any digital communications system is the following (4). It is used to get normalized signal-to-noise ratio $E_b/N_0$, where $E_b$ - average bit energy at the receiver input, $N_0/2$ - two-sided power spectral density of the noise at the receiver input.

$$\frac{E_b}{N_0} = \frac{S}{N} \cdot \frac{W}{R}, \qquad (4)$$

where  $S$ - average signal power at the receiver input,
$N$ - average noise power at the receiver input,
$W$ - bandwidth,
$R$ - data rate.

$$R = \frac{k}{T}, \qquad (5)$$

where $T$ - duration of symbol interval (0.3715 second)

$$\begin{array}{l} M = 64, \\ k = \log_2(M) = 6 \end{array} \qquad (6)$$

Bit error rate, given symbol error rate, can be expressed as follows [5]:

$$\frac{P_b}{P_e} = \frac{M/2}{M-1}. \quad (7)$$

Using (4) – (7) we found the dependence of bit error rate against the normalized signal-to-noise ratio (Fig. 10).

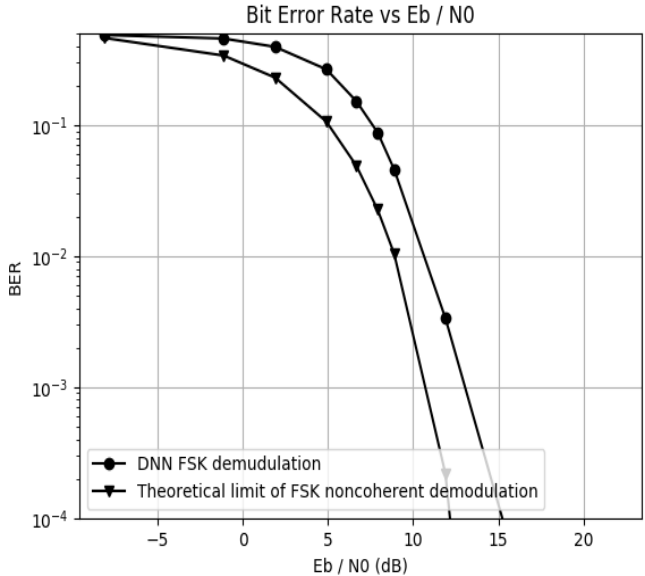

Fig. 10. Demodulation Bit Error Rate against normalized signal-to-noise ratio.

Also, the theoretical limit of bit error performance for non-coherent demodulation of orthogonal signals is presented in the Fig. 10 for comparison. Bit error probability for this case can be written as follows [5]:

$$P_b = \frac{1}{2} \exp\left(-\frac{1}{2} \frac{E_b}{N_0}\right). \quad (8)$$

BER performance (noise immunity) of the described data transmission method is less than theoretical performance limit of non-coherent demodulation of orthogonal MFSK signals for less than 2 dB at bit error probability level of $10^{-2}$.

Also we conducted the research of BER performance using combination of AWGN, narrow band sinusoidal interference (which is 20% of the AWGN power), and pulse interference (which is 10% of the AWGN power) at the inference time. The preliminary results show that the noise immunity is not affected more than for 0.45 dB at bit error probability level of $10^{-2}$.

## VIII. DISCUSSION

The overall purpose of the study was to prove the possibility of efficient demodulation weak MFSK signals using supervised deep machine learning. Our main finding suggests that the use of simple dense neural network has acceptable outcome. Such computational structures are easy to implement with modern microcontrollers or one-board microcomputers and software frameworks within the software part of Software Defined Radio (SDR) equipment. Software defined demodulation is performed by specialized computer programme that is used to recover the information content from the modulated carrier wave. Software defined demodulation approach has evolved to replace the traditional analogue or digital hardware counterpart. As stated above, the interference immunity of the data link is high enough and approaches close to the theoretical limit of non-coherent demodulation. Important advantage is ability to permanent retrain on new data. This allows to adapt to new conditions, signal and noise characteristics. Authors consider the approach as promising for data exchange in lattice based crypto systems [33]. The Jupyter notebooks with the experiments are open-sourced and can be downloaded from the GitHub repository [34].

## IX. SUGGESTIONS FOR FURTHER RESEARCH

One concern about the findings of noise immunity was the limitation to AWGN interference only. Another limitation was that artificially synthesized data do not reflect real complexity of the signals and interferences. Thus, it might be recommended to train further models on real data. Also, please note that we used fully connected layers in the neural network design. The use of convolutional layers or recurrent architectures possibly can be adapted flexibly, which will be our future work.

## X. CONCLUSION

The interference immunity of JT65A communication protocol as bit error probability against normalized signal-to-noise ratio of the data exchange by MFSK signals with ML based demodulation has been obtained for the first time. It has been proved that the interference immunity is about 2 dB less than the theoretical limit of non-coherent demodulation of orthogonal MFSK signals.